\documentclass[epj]{svjour}
\usepackage[final]{graphicx}
\usepackage{amsmath}
\usepackage{amssymb}
\begin{document}
\title{Optimal synchronization on strongly connected directed networks}
\subtitle{}
\author{Markus Brede\inst{1} 
\thanks{\emph{e-mail:} Markus.Brede@Csiro.au  }%
}                     
\offprints{}          
\institute{CSIRO Marine and Atmospheric Research, CSIRO Centre for Complex System Science, F C Pye Laboratory - GPO Box 3023, Clunies Ross Street
Canberra ACT 2601, Australia}
\date{Received: date / Revised version: date}
%
\abstract{In this paper we construct and analyse strongly connected sparse directed networks with an enhanced propensity for synchronization (PFS). Two types of PFS-enhanced networks are considered: (i) an eigenratio minimizing ensemble with non-vanishing complex parts of the spectrum and (ii) a class of networks with real spectrum but slightly larger eigenratios than (i). We relate the superior PFSs to a strongly skewed out-degree distribution, the density of double links and a hierarchical periphery-core organization. Ensembles (i) and (ii) are found to differ in the density of double links and the particular organization of the core and the periphery-core linkage.  
\PACS{
      {89.75.-k}{Complex systems.}	 \and
      {05.45.Xt}{Synchronization; coupled oscillators.}   \and
      {89.75.Fb}{Structures and organization of complex systems.}
     } 
} 
\maketitle
\section{Introduction}
Problems of the synchronization of coupled oscillators occur in various contexts. Relevant applications are found in ecology, neuroscience, biology and engineering, but also in the social sciences, e.g., in problems of consensus formation. A recent summary of relevant applications can be found in \cite{Arenas}.

Since the realization that the couplings of the individual elements in many of the above real-world systems are very heterogeneous and differ often markedly from spatial grids and unstructured random graphs  (cf. \cite{Barabasi,Newman,Bocc}), understanding synchronization on complex networks has attracted considerable attention in the recent literature. A good review of most of the recent developments in the field can be found in \cite{Arenas}. One of the important challenges is to understand the influence of network structure on the dynamics of synchronization. While various studies investigate properties of the synchronization transition of non-identical Kuramoto oscillators, much insight has also been gained by studying the linear stability of the synchronization manifold of identical oscillator systems via the master stability function approach \cite{Pecora}. In this work we follow the latter approach, employing the general eigenvalue criterion of \cite{Pecora} to classify a network's propensity for synchronization (PFS). This eigenvalue criterion is derived from a linear stability analysis of the completely synchronized state, see below. 

So far, with few exceptions as e.g. \cite{Timme,MB1,MB2} the literature has mainly focussed on oscillator systems coupled by undirected networks. Various studies have attempted to link network characteristics such as clustering \cite{Motter1,Zhao1,MB0,MB3}, degree mixing \cite{Motter1,Bernardo}, pathlength \cite{SW,Nishi} or betweenness centrality \cite{Nishi,Hong,Zhao1} to the PFS. More detailed analyses, however, show that such a characterization is at most meaningful in a statistical sense, since synchronization appears to be determined by `fine' details of the organization of the coupling network \cite{Atay}. Thus, linking synchronization properties to statistical network measures often provides only a rule of thumb.

Synchronization on directed networks has been indirectly addressed via introducing weighing schemes on networks, cf. \cite{Motter2,Hwang1,Chavez}. As main lessons this research shows that homogeneity in the in-signals of nodes as well as a top-down driven coupling structure, where coupling mainly occurs from hub nodes towards non-hub nodes facilitate synchronization \cite{Hwang1}.  However, notwithstanding these many results, insights about how the particular arrangement of links in directed graphs (with the many more degrees of freedom compared to undirected graphs) can influence synchronization properties are still rare.

Thus, in this paper, we address the issue of synchronization on directed unweighted networks. In this context Ref. \cite{Hwang1} makes it clear that directed networks with optimal PFS are directed trees. Synchronization on directed trees, however, can hardly be understood as a collective phenomenon, but is rather a top-down driven process. Moreover, most real-world networks on which synchronization plays a role are (i) inherently directed and (ii) strongly connected or at least have large non-trivial strongly connected components. Hence, in this paper we study strongly connected networks that display optimal PFSs.  The requirement of strong connectedness is also the reason why we limit the study to unweighted networks.

Using a numerical optimization scheme, we describe the structure of eigenratio-optimized strongly connected directed networks. As such our approach is similar to the one of \cite{Donetti}, where optimization is used to study optimal synchronization on undirected networks. Optimization techniques have already been used to understand the organization of complex networks in a variety of other contexts, e.g. \cite{Sole,Banvar,Variano,DN}, and we build on this body of knowledge here.

We find that directed strongly connected networks with optimal PFS have an hierarchical organization, i.e. they are composed of different subsets of nodes. While our finding that an optimal PFS requires the presence of connected hub and non-hub nodes and a hierarchical organization such that the former predominantly drive the latter confirms previous findings \cite{Hwang1,Chavez}, we also demonstrate additional structural requirements related to the details of the component organization and the emergence of a `super-hub' and its connections to the rest of the network. The latter, indeed, is a `fine' detail of the network organization not captured by statistical network properties such as pathlengths, motif densities or degree distributions.

In the following we give a brief description of the general problem and the master stability function approach that leads to the eigenratio criterion to define a network's PFS. We then motivate the numerical optimization scheme, give a brief discussion of the dynamics of Roesler oscillators on the optimal networks, and continue with the structural analysis of the networks with optimal PFS. The last section summarizes and discusses the main results.

\subsection{The MSF Approach}
\label{msf}
A general class of synchronization problems can be described in the following way. Consider a system composed of $N$ identical oscillators. Without coupling the individual dynamics of each oscillator is given by
\begin{align}
\dot{x}_i=F(x_i),
\end{align}
where $x$ is the state vector in some $m$ dimensional space and $F$ gives the individual dynamics of the oscillators without coupling. Now assume that the individual oscillators are coupled by
\begin{align}
\label{Dyn}
\dot{x}_i &= F(x_i)+\sigma \sum_{j=1}^N A_{ij} (H [x_j]-H[x_i])\\
          &= F(x_i)+\sigma \sum_{j=1}^N G_{ij} H [x_j]
\end{align}

In Eq. (\ref{Dyn}), $H$ is an output function that determines which information about the state of $i$ is transmitted to $j$, $\sigma$ the coupling strength and $A$ with $A_{ij}=1$ if $j$ is linked to $i$ and $A_{ij}=0$ otherwise is the adjacency matrix of the coupling network. The second line of Eq. (\ref{Dyn}) rewrites the dynamics by introducing the Laplacian matrix of the graph defined by its adjacency matrix $A$. Since the effective coupling matrix $G$ has zero row-sum, a completely synchronized state $x_1(t)=...=x_N(t)=s(t)$ can exist.

 One then analyzes the linear stability of small perturbations $\delta x_i$ around the synchronized state and obtains
\begin{align}
\label{Stab}
 \delta \dot{x}_i =   D F (s) \delta x_i + \sigma DH (s) \sum_{j=1}^N G_{ij} \delta x_j,
\end{align}
with $DF(s)$ and $DH(s)$ being the Jacobian matrices of $F$ and $H$ at $s$. As argued in \cite{Pecora}, Eq. (\ref{Stab}) can be diagonalized into $N$ decoupled modes by projecting $\delta x$ into the space spanned by the eigenvectors of $G$, yielding
\begin{align}
\label{MSF}
 \dot \eta_i= [DF (s)- \sigma \lambda_i DH (s)] \eta_l, l=1,...,N.
\end{align}
The eigenmode corresponding to $\lambda=0$ describes perturbations parallel to the synchronization manifold, whereas the remaining $N-1$ transversal eigenmodes are required to decay for the fully synchronized state to be stable. Equation (\ref{MSF}), which is essentially the same for all the non-zero modes, defines the master stability function. Introducing $\alpha_i=\sigma \lambda_i $ the evolution of small perturbations $\eta$ is given by the largest Lyapunov exponent of the r.h.s. of (\ref{MSF}), the so-called master stability function (MSF) \cite{Pecora}. It has been shown that for many oscillator types this eigenvalue is smaller than zero only in a limited region of the complex plane \cite{Fink}. 

Let us order the real parts of the eigenvalues of the coupling matrix $G$ in the following way $0=\lambda^r_1\leq \lambda^r_2\leq...\leq \lambda^r_N$. Since the coupling network is strongly connected we have $\lambda^r_2>0$. Moreover, since the entries of $G$ are positive real numbers, complex eigenvalues occur in pairs of complex conjugates. Hence, if the stability region of (\ref{MSF}) is bounded,  the eigenratio $r=\lambda^r_N/\lambda^2_2$ and the largest magnitude of the complex parts of the eigenvalues $c=\max_{j>1} |\lambda^i_j|$ of $G$ must be small. In this sense, the eigenratio $r$ and the maximum spread of the complex parts $c$ are measures for a network's PFS. Should the stability region of (\ref{MSF}) be unbounded in the real domain, the eigenratio $r$ is to be replaced by $1/\lambda^r_2$, i.e. configurations with maximum $\lambda^r_2$ realize optimal PFS. Our following analysis and comparison between networks PFS's will be based on the assumption of a bounded stability region of the MSF, i.e. the first criteria.

\subsection{The optimization procedure}
\label{OptProc}
\begin{figure}
\begin{center}
\includegraphics [width=.45\textwidth]{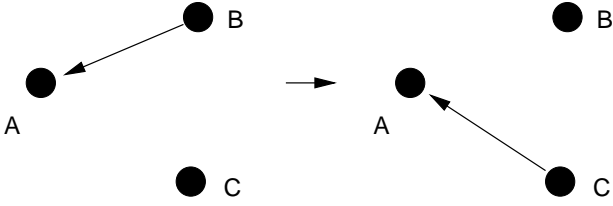}
\caption{Rewiring procedure that leaves the in-degree sequence unchanged.}
\label{F1}
\end{center}
\end{figure}

The overall aim of the paper is to construct strongly connected directed networks with an enhanced PFS by an optimization procedure. Generally, we employ the following algorithm, similar to, e.g. \cite{MB1}: (i) start with a random network configuration, (ii) suggest random rewirings that leave the number of links constant and do not disconnect the strong component and (iii) evaluate the PFS of the new configuration as characterized by the eigenration $r$ and the maximum absolute value of the complex parts of the eigenvalues $c$. Modified configurations are accepted if they have a PFS superior or equal to the previous configuration and rejected otherwise. The procedure is iterated till no improvement in the PFS can be obtained any more for a given number of rewiring attempts that is proportional to the number of links in the network. In the above procedure, we generally start by suggesting multiple link rewirings at a time, this `exploration environment' being gradually narrowed down as the optimization progresses.

In agreement with several studies in the literature \cite{Motter2,Hwang1,Chavez,Donetti} preliminary experiments with the above procedure demonstrated that homogeneity in the in-signals (i.e. in-degree homogeneity) has a predominant influence on the eigenratio, without affecting the maximum complex part $c$ in a systematic fashion. Hence, we developed a refined optimization procedure that restricts itself to in-degree regular graphs. Rewiring suggestions are arranged in the following way which leaves the in-degree sequence of the graph unchanged. A node $A$ and one of its out-neighbours $B$ are chosen at random. Then, a new out-neighbour $C$ for $A$ (which is not $A$ itself and to which $A$ is not already connected) is chosen at random, provided that the rewiring $A\to B$ to $A\to C$ would not disconnect the strong component. The procedure is illustrated in Fig. \ref{F1}.

\begin{figure}
\begin{center}
\includegraphics [width=.45\textwidth]{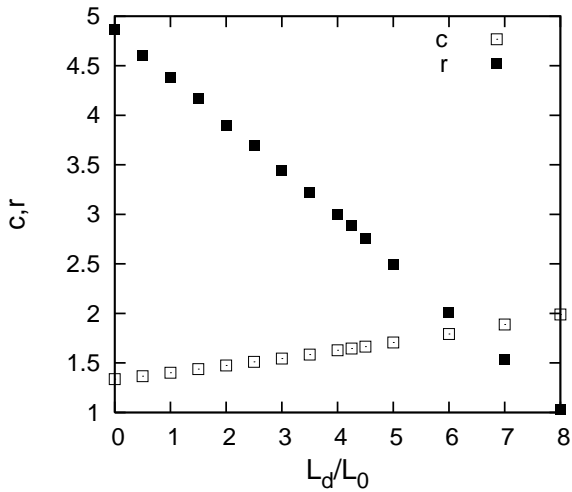}
\caption{Dependence of the extent of the complex part of the spectrum $c$ and the eigenratio $r$ of the Laplacian matrix on the number of double links in the network.}
\label{F2}
\end{center}
\end{figure}

Next, arguments about the spectra of large random symmetric/antisymmetric matrices indicate that a positive/negative correlation between transposed matrix elements reduces/enlarges the extent $c=\max_{j>1} |\lambda^i_j|$ of the complex part of the spectrum \cite{Sommers}. For the special case of positive binary coupling networks of interest here this suggests a relationship between the link reciprocity and the quantity $c$, i.e. an enhanced number of double links reduces the complex part of the spectrum. This is not surprising, as one  network configuration minimizing $c$ are fully bidirectional networks corresponding to symmetric Laplacian matrices which have a real spectrum. In fact, experiments constructing networks that minimize $c$ for the Laplacian matrix of the coupling network show that symmetric coupling proves to be the attractor of the optimization in all cases (data not shown). Minimizing $r$ for configurations with $c=0$ then reproduces the known results for synchrony-optimized networks for undirected networks \cite{Donetti}, which the authors termed `entangled nets'. These results, however, do not make use of the additional degrees of freedom for network organization that are available in directed networks.

On the other hand, numerical experiments with in-degree regular random coupling networks with tuned link reciprocities also suggest a strong negative correlation between the eigenratio and the link reciprocity. To demonstrate this relationship we constructed ensembles of in-degree regular networks of $N=100$ nodes with a large link density $\langle k_\text{in}\rangle=\langle k_\text{out}\rangle=20$ and systematically tuned link reciprocity, measured by the normalized number of double links $L_d$. For the parameter range investigated this tuning can be achieved without a significant change in the diameter and average pathlength of the network. Figure \ref{F2} gives the dependence of both the quantities $r$ and $c$ on the normalized density of double links $L_d/L_0$, where $L_0$ is the number of double links expected in a random directed graph. The numerical data exhibit an almost linear decrease in $r$ and a linear increase in $c$ when growing the density of double links. A more detailed investigation reveals that in fact both the smallest non-trivial real part of any eigenvalue $\lambda^r_2$ and the largest real part $\lambda^r_N$ of $G$ contribute to the dependence of the eigenratio on $L_d$. That is, while $\lambda^r_N$ systematically decreases with $L_d$, $\lambda^r_2$ is found to increase. Similar relationships between $c$ and $r$ and $L_d$ have been obtained for in-degree regular graphs with more/less skewed out-degree distributions and the results are also robust when the link density is changed.

The above arguments suggest that a simultaneous maximization of $r$ and minimization of $c$ are not possible. In fact, the detailed requirements on $r$ and $c$ are determined by the detailed structure of the stability region of the MSF. However, even though a simultaneous maximization of $r$ and minimization of $c$ proved impossible, network structures with a strongly enhanced PFS can be constructed by maximizing $r$, while requiring $c$ to remain bounded. Hence, in the following, we optimize the eigenratio of the coupling networks, but only accept rewired configurations for which the complex part of the spectrum is bounded $c\leq c^*$, i.e. assume that the stability region of the MSF is such that minimization of the eigenratio is the dominant contributor to the stability of the synchronized state. 

\begin{figure}
\begin{center}
\includegraphics [width=.45\textwidth]{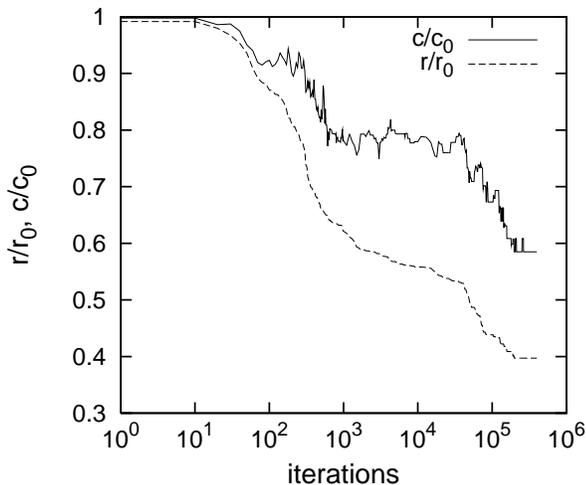}
\caption{Evolution of the normalized eigenratio $r/r_0$ and maximum complex part $c/c_0$ during the optimization. The quantities $r$ and $c$ are normalized by the $r_0$ and $c_0$ at iteration $0$ of the optimization (i.e. by the average values for in-degree regular random graphs). The data are obtained from 10 optimizations of networks of size $N=100$ with $\langle k_\text{in}\rangle=\langle k_\text{out}\rangle=3.$}
\label{F3}
\end{center}
\end{figure}

Figure \ref{F3} shows the evolution of $r$ and $c$ in the course of such an optimization. Two observations stand out: (i) even though $r$ and $c$ are not simultaneously minimized, the eigenratio $r$ can be minimized while $c$ remains bounded and in fact also decreases by roughly one third and (ii) a very substantial reduction of $r$ in comparison to the undirected optimum configuration (i.e. `entangled networks') is possible. In the following section we proceed by analyzing the networks with an enhanced PFS obtained in this way.

\section{Results}

\begin{figure}
\begin{center}
\includegraphics [width=.49\textwidth]{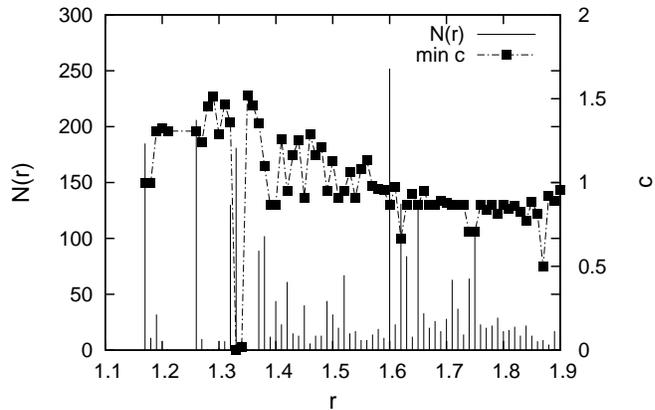}
\caption{Histogram of numbers of networks constructed with eigenratio $r$ (bin size $.01$). The dashed line gives the minimum of the extents of the complex part of the spectrum for each bin, the points are connected to guide the eye (y2 scale). We pick the two outstanding ensembles for further analysis: (a) networks that minimize the eigenratio ($r\approx 1.17$ and $c=1.$) and (b) networks that minimize $c$ while having a very low eigenratio (the group with $r\approx 1.34$ and $c\approx 0.$).   }
\label{F4}
\end{center}
\end{figure}

For the following analysis about 3000 sparse directed networks of size $N=100$ with $\langle L\rangle=300$ links have been optimized for an enhanced PFS according to the procedure outlined in the previous section. It turns out that the `fitness' landscape defined by the networks' PFS is riddled by many separate `deep' basins of attractions, such that the optimization frequently gets stuck in local optima. Figure \ref{F4}, which gives a histogram for the eigenratios of the constructed networks gives an overview over the optimization results. While the left y-axis (impulses) shows the numbers of networks that could be constructed with a certain eigenratio, the opposite y-axis (solid squares connected by a dashed line) indicates the minimum extent of the complex part of the spectra over all networks with the same eigenratio. Analyzing the data displayed in Fig. \ref{F4} two ensembles of networks stand out: (a) The configurations for which the smallest eigenratios are obtained ($r\approx 1.17$). We consider the subset of networks which have the smallest extent of the complex part of the spectrum in this group, i.e. $c=1.0$. In the following we refer to this group as $r$-optimized networks. (b) The configurations for which the spectrum is real ($c\approx 0$). We consider the subset of networks with the smallest eigenratio in this group, i.e. configurations for which $r\approx 1.34$. In the remainder of the paper this ensemble is referred to as $r/c$-optimized networks. Ensemble (a) is of interest because it represents the network with optimal PFS if the extent of the real domain of the stability region  of (\ref{MSF}) is the dominant consideration. Ensemble (b) represents networks with optimal PFS in all cases where both the real and the complex part have to be taken into consideration. The number of evolved PFS-enhanced network was chosen such that in both groups, the $r$- and $r/c$-optimized ensembles, $100$ networks could be obtained.

Albeit not reported in detail here, we have also evolved optimal sparse networks of different connectivities and sizes. It is important to note that the results described below were found to be essentially robust in all examined situations, i.e. we always found a structurally similar $r$-optimal ensemble of networks minimizing $r$ (but having distinctly non-zero $c$) and an $r/c$-optimal ensemble minimizing $c$ (but having a value of $r$ distinctly inferior to the $r$-optimal ensemble).

\subsection{Dynamics}

Before proceeding with an analysis of the evolved networks, we pause to demonstrate that the above procedure, which is based on an investigation of the linear regime close to the fully synchronized state, does indeed result in networks on which a higher degree of synchronization can be achieved. Even though the linear stability analysis of subsection \ref{msf} strictly applies only for systems of identical oscillators, similar to, e.g., \cite{Chavez}, we find the PSF to be a good indicator of a `synchronizability' for systems of non-identical oscillators as well.

As an example system for Eq. (\ref{Dyn}) let us consider a system of coupled Roesler oscillators with coordinates $(x_i,y_i,z_i)$ for which
\begin{align}
 F(x_i)= \begin{pmatrix} -\omega_i y_i -z_i\\ \omega_i x_i+0.165 y_i\\ 0.2+z_i (x_i-10) \end{pmatrix}.
\end{align}
Let the inner coupling be realized by the function $H(x)=x$ and let the natural frequencies $\omega_i$ be drawn from a uniform distribution over the interval $[0.98,1.02]$.

To proceed, we have numerically integrated Eq. (\ref{Dyn}) for the above system using a 4th order Runge-Kutta  method. Phase synchronization in this system can be measured by an order parameter
\begin{align}
 r_\phi (t)= 1/N \sum_j \exp(i \phi_j (t)),
\end{align}
where the phases can be defined via
\begin{align}
\phi_j(t)=\arctan (y_j(t)/x_j(t)).
\end{align}

On the other hand, 
\begin{align}
 \nonumber
 \delta_x (t) &= 1/\sqrt N \times \\
              & \left (\sum_{j} (x_j-\langle x\rangle)^2+(y_j-\langle y\rangle)^2+(z_j-\langle \rangle)^2 \right)^{1/2},
\end{align}
with $\langle x\rangle=1/N\sum_j x_j$, $\langle y\rangle=1/N\sum_j y_j$, and $\langle z\rangle=1/N\sum_j z_j$ is a measure for the synchronization error. To obtain values between zero and one we further normalize $\delta_x(t)$ by its maximum value and introduce $r_\delta (t)=1-\delta(t)/\delta_\text{max}$. Averaging $r_\phi (t)$ and $r_\delta (t)$ over a suitable interval of time after relaxation, over different randomly assigned choices of native frequencies $\omega_i$ and over the optimized networks generated via optimizing the PSF as described above leaves us with two order parameters for synchronization, i.e. $r_\phi$ for phase synchronization and $r_\delta$ for the desynchronization error. In both cases we have $r=0$ if the oscillators are desynchronized and $r=1$ when they are fully synchronized.

\begin{figure}
\begin{center}
\includegraphics [width=.4\textwidth]{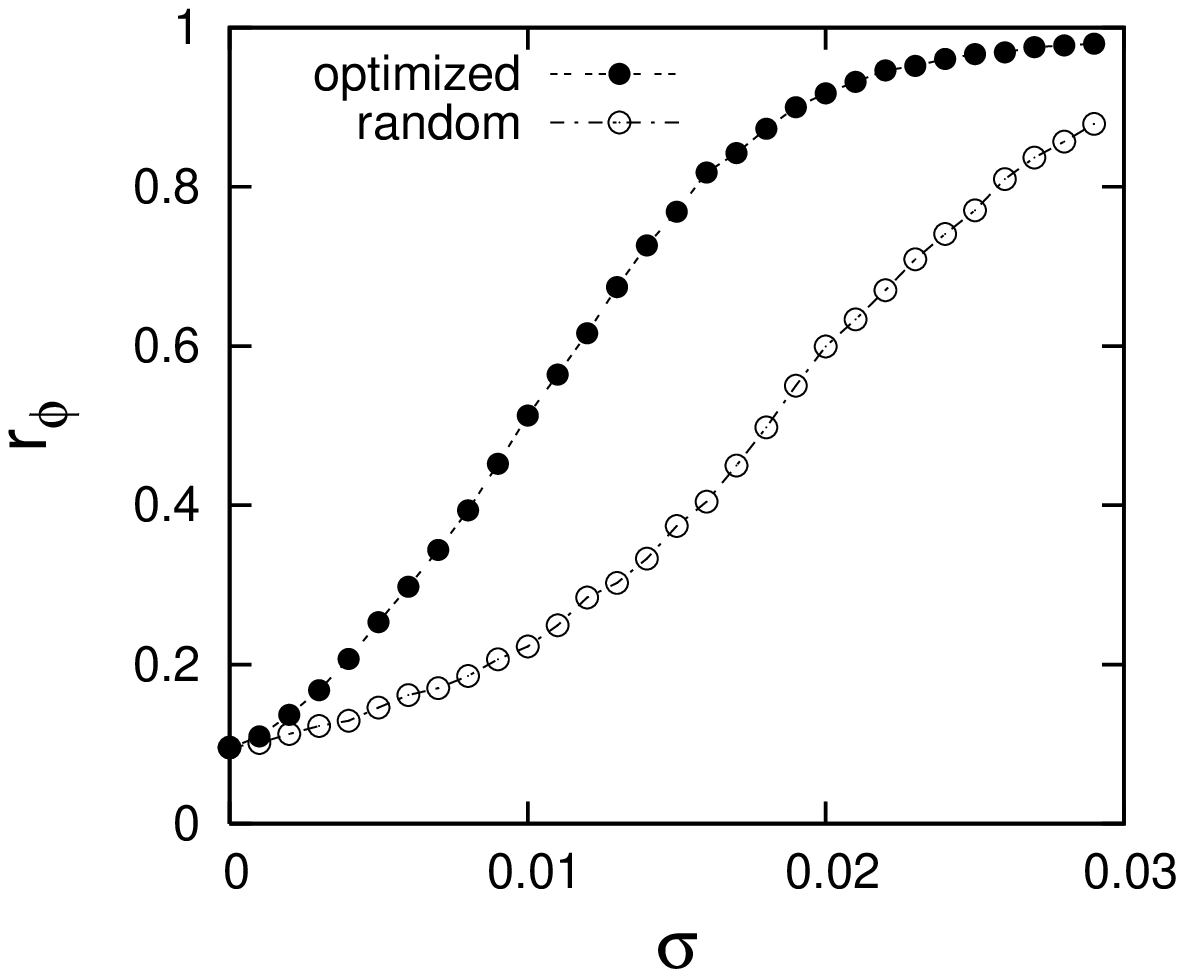}
\includegraphics [width=.4\textwidth]{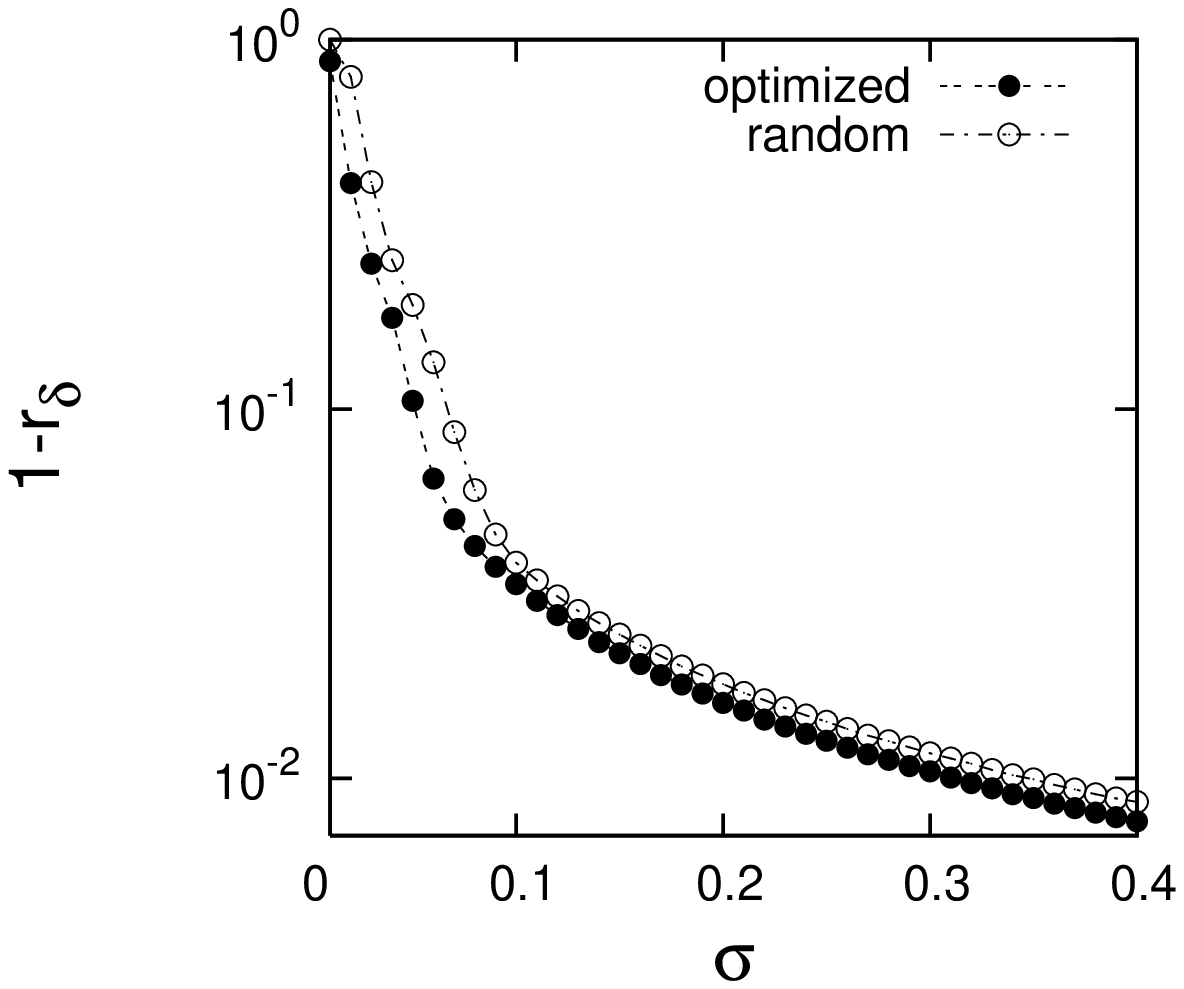}
\caption{Dependence of the order parameters $r_\phi$ (top) and $r_\delta$ (bottom) on the coupling strength. The figure compares the synchronization transition for $N=100$ Roesler oscillators coupled by the optimized networks (full circles) and by random in-degree homogeneous networks of the same connectivity (open circles). Note the different ranges of coupling strengths and the logarithmic y-scale at the bottom figure.}
\label{FRA}
\end{center}
\end{figure}

By plotting the dependence of $r_\phi$ and $r_\delta$ on the coupling strength Figure \ref{FRA} compares the synchronization behaviour of random in-degree homogenous networks and PSF-optimized network ensembles. The latter clearly display an earlier transition to phase synchronization, but also a generally lower desynchronization error, supporting the point that the linear analysis is to some degree indicative of a network's synchronization behaviour.

\subsection{Out-degree variance, pathlengths and motif densities}

\begin{figure}
\begin{center}
\includegraphics [width=.49\textwidth]{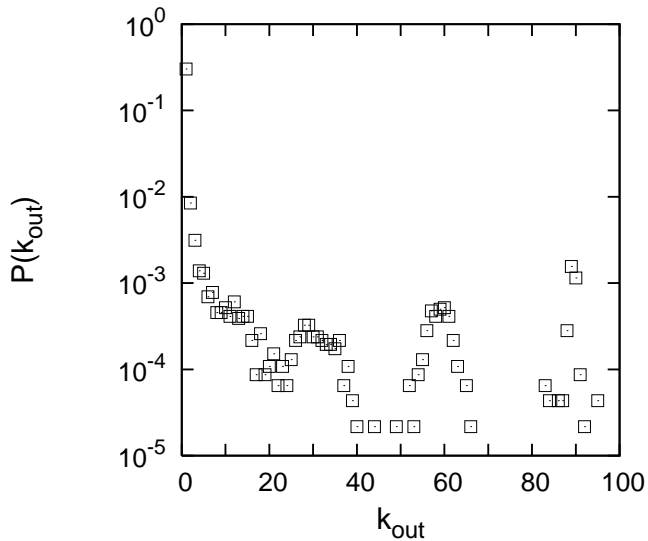}
\caption{Degree distributions of the optimized networks (groups (a) and (b) combined).}
\label{F5}
\end{center}
\end{figure}

Before investigating the ensembles (a) and (b) separately in more detail, it is worth noting that both network ensembles are marked by very strong out-degree heterogeneity and have similar out-degree distributions, cf. Fig. \ref{F5}. In fact, the statistics for the out-degree distributions indicate that the evolved networks essentially consist of a group of periphery nodes (which are mostly leaves or have very small degree) and a set of hub nodes that form the core of the network. Gaps in the out-degree hint to a hierarchical organization of the hub-nodes. Closer inspection shows that a typical optimized network comprises one `super-hub', marked by a distinctly larger out-degree than all other nodes. Three to four further hubs with out-degrees corresponding to further maxima of the out-degree distribution can be distinguished, such that the core of the networks typically consists of four to five nodes. More formally, the core may be defined as the set of nodes $C_k=\{i| k_\text{out}(i)\geq k\}$ (and the set of links connecting nodes of the core) and the periphery by $P_k=\{i|k_\text{out}(i)<k\}$ (and the set of links connecting periphery nodes). Analyzing the out-degree distribution a cut-off of $k=5$ appears a sensible choice \cite{Remark}. 

Proceeding in our analysis, we next analyzed frequency counts of motifs for the network ensembles. The motif counts thus obtained are compared to random expectations, which are calculated from randomized network ensembles with the same degree distributions as the original ensembles. Most prominently, statistically significant differences in the densities of double links are found, cf. the summary of network statistics in Tab. \ref{tab1}. As expected from the argument explained in subsection \ref{OptProc} there are only very few double links in the $r$-optimized networks. In contrast, the $r/c$-optimized networks show a slightly increased density of double links. In fact, experimentally perturbing the structure of the optimized networks by artificially increasing/decreasing the number and location of the double links proves that their number and arrangement are significant contributors to the enhanced PFSs. For instance, rewiring an $r$-optimal network such that only one double link between periphery nodes is added, typically reduces the eigenration to about the mean value of the ensemble of in-degree regular random graphs.

In all evolved networks the double links are found to be connected to at least one core node. Occasionally occuring double links for $r$-optimized networks typically connect core nodes (excluding the super-hub), while double links in the $r/c$-optimized networks either establish links between core nodes or links between a periphery node and a core node. However, comparing randomized networks which obey these placement constraints ($r=2.49,c=1.16$) and such that do not ($r=2.53$, $c=1.19$) suggests that the explanatory power of these rules is not sufficient to explain the difference in the PFS of the evolved and networks and random networks.

An analysis of various triangle counts and higher order motifs also showed no statistically significant differences to random networks with the same in- and out-degree distributions. This finding is in itself of interest, as some previous studies have hinted to the existence of a correlation between the synchronizability of a motif and its frequency count in several real world networks \cite{Moreno1,Moreno2}. One may thus surmise that such enhanced densities of more synchronizable motifs also hints to an enhanced PFS of the system as a whole. Together with an argument advanced in \cite{MB2} our finding here, that optimized networks do not appear to have larger than random densities of any motif apart from double links, falsifies this hypothesis.

An analysis of the distance statistics of the constructed networks shows that they are considerably larger than the in-degree homogeneous random network with which we seeded the optimization, but also considerably smaller than randomized networks with the same in- and out-degree distributions than the constructed networks, see Tab. \ref{tab1} which summarizes various network statistics. The effect is much reduced for the $r/c$-optimal networks, which suggests that the differences in distance is caused by the different densities of double links.

Altogether, the analysis of the standard network properties suggests that (i) the out-degree variance is a strong predictor of the PFS and (ii) that the density of double links systematically relates to the PFS. Both network characteristics, however, are in themselves not sufficient to explain the enhanced PFS of the evolved networks.

\begin{table*}
\caption{Overview about properties of PFS-optimized graphs ($N=100$, $L=300$)}
\label{tab1} 
\begin{center}  
\begin{tabular}{lllllll}
\hline\noalign{\smallskip}
ensemble & $\langle r\rangle$ & $\langle c\rangle$ & $\langle \sigma^2_{k_\text{out}}\rangle$ & $\langle L_\text{d}\rangle$ & $\langle l\rangle$ & $\langle l_\text{max}\rangle$\\
\noalign{\smallskip}\hline\noalign{\smallskip}
undirected regular random & $27.3$ & $0.0$  & $0.0$  & $150.0$     & $4.84$ & $8.8$\\
undirected, `entangled'   & $12.5$ & $0.0$  & $0.0$  & $150.0$     & $4.44$ & $7.0$\\
in-degree regular random  & $3.47$ & $1.67$ & $2.4$  & $4.5\pm .3$ & $4.0$  & $8.4$\\
$r$-optimal directed      & $1.17$ & $1.00$ & $118.$ & $1.1\pm .2$ & $6.2$  & $13.9$\\
$r/c$-optimal directed    & $1.34$ & $0.0 $ & $117.4$& $6.0$       & $8.8$  & $19.3$\\
$r$-opt., randomized      & $2.37$ & $1.25$ & $118.$ & $4.6\pm.6 $ & $9.5$  & $23.6$\\
$r/c$-opt., randomized    & $2.39$ & $1.25$ & $117.0$& $4.8\pm .5$ & $9.6$  & $24.3$\\
$r$-opt., randomized,cd   & $2.18$ & $1.56$ & $118.$ & $1.1\pm .2$ & $8.6$  & $22.2$\\
$r/c$-opt., randomized,cd & $2.53$ & $1.19$ & $117.0$& $6.0$       & $9.9$  & $24.9$ \\
\noalign{\smallskip}\hline
\end{tabular}
\end{center}
\end{table*}

\subsection{Organization of the cores}
In the following section we analyze the organization of the cores $C_k$, $k=5$, of the optimized networks. These are then compared to randomized networks with the same degree distribution, i.e. the same size of the core. For $k=5$ the average size of the core is $|P_5|=4.8$, i.e. typically only about five nodes have an out-degree larger than $k=5$.

To elicit the structure of the core, the following quantities are of interest. First, the average number of core-links $L(P_k)$ characterizes the cohesiveness of the core. Next, the maximum out-component
\begin{align}
 |C_\text{out}^\text{max}| = \max_{i\in P_k} |C_\text{out} (i)|,
\end{align}
where $C_\text{out}(i)$ gives the maximum set of core nodes that can be influenced by $i$, decides whether the whole core can in principle be directed by one core node. This information is refined by calculating the maximum size of the strong component
\begin{align}
|SCC(P_k)|=\max_{i\in P_k} |SCC(i)|,
\end{align}
a measure for the mutual influence between core nodes. A more detailed investigation of individual core arrangements suggests to also define a quantity $|C_{mo}|$,  the average size of the in-component of nodes with maximum out-component. The relevant statistics for the cores of the optimized network ensembles and their randomized counterparts are detailed in Tab. \ref{tab2}.

\begin{table}
\caption{Analysis of the cores}
\label{tab2} 
\begin{center}  
\begin{tabular}{lcccc}
\hline\noalign{\smallskip}
ensemble & $|SCC(P_k)|$ & $|C_\text{out}^\text{max}|$ & $|C_{mo}|$ & $L(P_k)$\\
\noalign{\smallskip}\hline\noalign{\smallskip}
$r$-opt.      & $2.4$        & $4.8$               & $1.1$    & $8.7$ \\
$r$-opt. (r)  & $2.6$        & $4.8$               & $2.5$    & $7.2$ \\ 
$r/c$-opt.    & $3.7$        & $4.8$               & $3.3$    & $10.2$\\
$r/c$-opt. (r)& $3.0$        & $4.8$               & $3.0$    & $8.0$\\
\noalign{\smallskip}\hline
\end{tabular}
\end{center}
\end{table}

\subsubsection{$r$-optimized networks}
Cores of the $r$-optimal network ensembles are found to be more densely linked than the randomized ensembles. However, the arrangement of links is different from random. Even though the maximum in-component comprises the whole core, a unique node, the `super-hub', that influences all other nodes in the core is found. This node is not influenced by any other node within the core, i.e. $|C_{mo}|\approx 1$. The strong component within the core is nevertheless almost equal to that of the randomized ensemble, and is typically comprised of the hubs of intermediate out-degree $20<k_\text{out}<70$, cf. also Fig. \ref{F5}.

It is interesting to note that the constraint of strong connectivity does not lead to an optimal network organization that is comprised of a strongly in itself interconnected core with minimized input from the periphery in this case, but rather to an organization that is controlled by a single super-hub, which receives all its input from the periphery. As we will see below, this input stems from the most distant nodes which distinguishes the super-hub from the other core nodes. They, in turn, receive all their input from other core nodes or nodes belonging to a generalized core.

\subsubsection{$r/c$-optimized networks}
The organization of the components within the $r/c$-optimized networks is essentially different from those of the $r$-optimized networks. Cores of $r/c$-optimized networks typically comprise a relatively large strong component, which also includes the super-hub. These nodes within the strong component are strongly interlinked, which contributes to the substantially larger than random link density in the cores. Closer inspection shows that the strong component mostly contains only the largest out-degree hub nodes, thus leaving one or two core nodes with relatively low out-degree without input into it. The statistics of periphery nodes influencing core nodes is also essentially different from that of the $r$-optimized networks. As explained in more detail below, the largest out-degree hubs (and thus the nodes in the strong component of the core) receive input from nodes that belong to the extended core, rather than from nodes in a greater distance as for the case of the $r$-optimal networks.

\subsection{Hierarchies}
\begin{figure}
\begin{center}
\includegraphics [width=.25\textwidth]{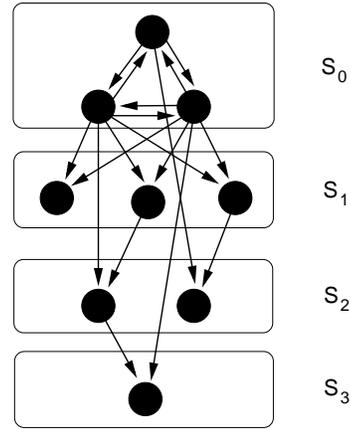}
\caption{Illustration of the definition of core sets for an example network, cf. text.}
\label{FH0}
\end{center}
\end{figure}

A detailed investigation of many individual optimized networks suggests that these networks are distinguished by a hierarchical organization of nodes according to the `quality' of the in-signals they receive. To capture this organization, we define a hierarchy of core sets $S_d$ in a recursive way. By definition $S_0=P_k$. Then, let $S_{d+1}$ be comprised of all nodes that receive input only from lower level core sets, i.e. from nodes of the set $\cup_{i=1}^d S_i$. The core cluster $S_\text{max}=\cup_i S_i$ is the set of nodes that can be reached that way.  All nodes in the core cluster $S_\text{max}$ are uniquely assigned to one set, say $S_d$, and are thus assigned a well-defined value of a core distance $d$. The maximum core distance defines the core diameter $d_\text{max}$ and the average core distance is obtained from
\begin{align}
 \langle d\rangle=1/|S_\text{max}| \sum_{i=1}^{d_\text{max}} i|S_i|.
\end{align}

An illustration of the concept is provided in Fig. \ref{FH0}. For the in-degree regular network displayed in the figure the core $S_0$ is defined by $C_3$, all nodes with out-degree larger are equal to three. Nodes in $S_1$ then exclusively get input from nodes in $S_0$, nodes in $S_1$ get some input from $S_0$, but also from $S_1$, etc. In the example the core cluster comprises the whole network and the core diameter is three.

It is important to note that nodes at a level $S_d$ cannot influence nodes of lower levels $S_i$, $0<1<d$. However, nodes from every level can have input into the core $S_0$ and in fact some of them must have input into the core to meet the constraint of strong connectivity for the overall network.

The interpretation of these core sets as a hierarchy of nodes with different qualities of input information is close at hand. Since more than half of the links in the network starts from core nodes $P_k$, one can argue that the network is predominantly driven by those nodes, such that the state variables of the core nodes will be closest to the synchronization manifold. States of nodes that receive input by core nodes without being perturbed by influences of other nodes will be the next closest, etc. 

Measuring the core-set and core-distance charcteristics of the evolved networks reveals their hierarchical organization. For $r$-optimized networks one finds $|S_\text{max}|=99.5\pm 1.$ and for $r/c$-optimized networks $|S_\text{max}|=95.7\pm 2$. These values compare to $|S_\text{max}|=85.3\pm 3.$ and  $|S_\text{max}|=84.\pm 3.$ for the respective randomized ensembles, i.e. in the optimized networks almost all nodes are part of the core cluster and are substantially larger than the expectations.

More interestingly, notwithstanding the larger core clusters, the average core distances of the evolved networks are substantially smaller than the random expectations. For the $r$-optimized networks one obtains $\langle d\rangle=2.6\pm.2$ (random expectation $\langle d\rangle=5.\pm.5$). Similarly one has  $\langle d\rangle=3.4\pm.2$ for the $r/c$-optimized networks (random expectation $\langle d\rangle=5.1\pm.6$). Smaller average core-distances also translate into smaller core-diameters, i.e.  $\langle d_\text{max}\rangle=11.8\pm 1 $ vs. $\langle d_\text{max}\rangle=22.3\pm 2$ and $\langle d_\text{max}\rangle=17.\pm 1 $ vs. $\langle d_\text{max}\rangle=24.3\pm 2 $ for the respective ensembles. These differences in the core-distances are substantially larger than the differences in the average pathlengths and diameters, cf. Tab. \ref{tab1} and suggest that the optimized networks optimize the quality of the input information, whereas avg. pathlengths and diameter can remain relatively large. In fact, a more detailed investigation shows that the large average distances and diameters in the optimized networks mainly arise from distances between pairs of periphery nodes, which constitute the bulk of the network.

\begin{figure}
\begin{center}
\includegraphics [width=.49\textwidth]{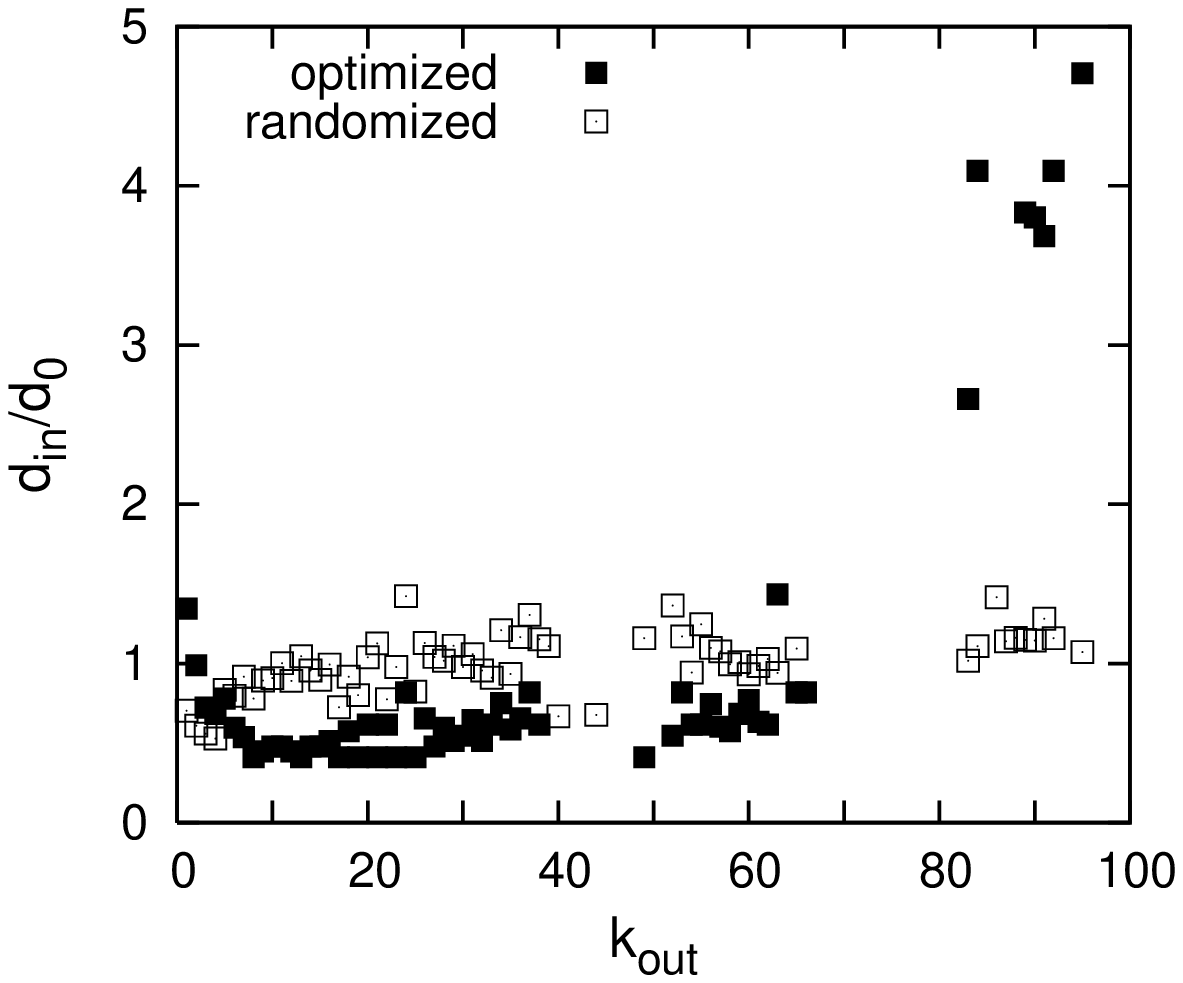}
\includegraphics [width=.49\textwidth]{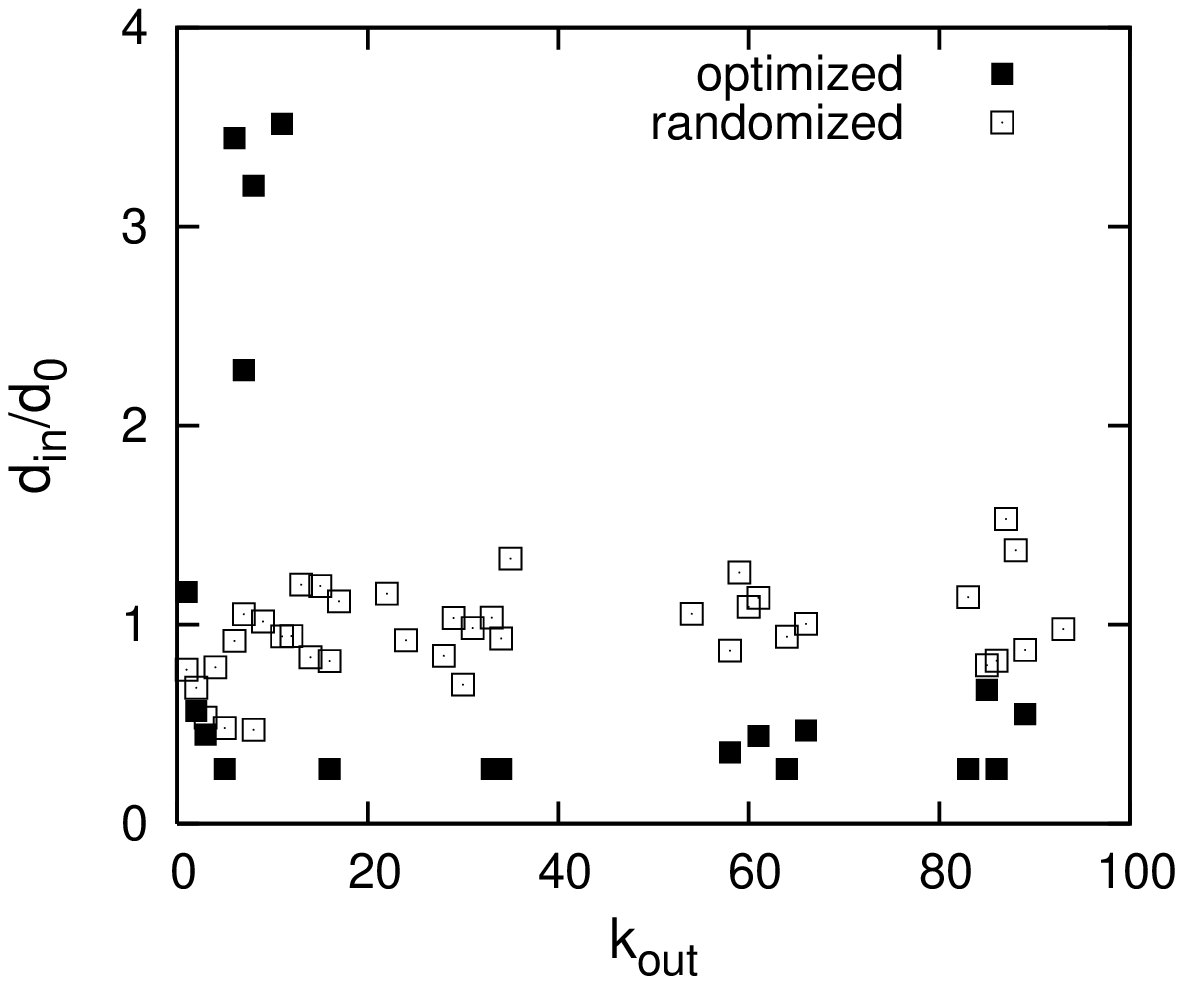}
\caption{Dependence of the avgerage level distance of inputs into a node on the node's out-degree for $r$-optimized networks (top) and $r/c$-optimized networks (bottom). }
\label{FH1}
\end{center}
\end{figure}

Further interesting information about the network organization can be obtained from an analysis of the core distance of nodes that have input into the core and the periphery. Figure \ref{FH1} details the dependence of the average core distance of non-core nodes that have input into a node on the node's out-degree, i.e.
\begin{align}
 d_\text{in}(i)=\frac{\sum_{j\in P_k} A_{ji} d(j)} {\sum_{j\in P_k} A_{ji}},
\end{align}
provided the node $i$ receives any input from nodes other than the core. The data are normalized by the average non-core input $d_0$ and compared to the expectation calculated by an evaluation of randomized networks. Hence ratios $d_\text{in}/d_0<1$  indicate input of nodes close to the core and $d_\text{in}/d_0>1$ input of nodes far away from the core.

We start with the analysis of the $r$-optimized networks. Whereas the expectation essentially shows no dependence of a node's non-core input on its out-degree a clear pattern is revealed for the optimized networks. The outstanding observation from Fig. \ref{FH1} is that the super-hubs, which as we have seen in the previous section receive only input from the periphery, receive this input from the periphery nodes with the largest core distance. In contrast, other core nodes typically receive input only from such non-core nodes that are already very close to the core, i.e. essentially from nodes belonging to $S_1$. In this sense one can consider the core nodes and the nodes in $S_1$ as a tightly interlinked extended core.

The organization of the cores of the $r/c$-optimal networks is again different. Here, particularly high-degree core nodes receive input from nodes in the core set $S_1$, thus again constituting an extended core. Feedback from the periphery to the core occurs via the lowly connected (and thus lower degree) core nodes. However, common to both network organizations is a tighly packed formation of a hierarchy of extended cores driven by a tightly interlinked core in the centre, which then gradually loses coherence as it extends towards the periphery.

\section{Conclusions}

Trying to fill the gap in knowledge about synchronization properties of directed graphs we have constructed and analyzed synchrony-optimized strongly connected directed in-degree regular networks. In our analysis a network's PFS is evaluated on the basis of the eigenvalue criterion derived in \cite{Pecora}. The choice of in-degree regular ensembles is motivated by the established result that heterogeneity in the in-signals is a main contributor to a network's PFS and also by the fact that many real-world networks  have a relatively narrow in-degree distribution. The constraint of strong connectivity was introduced because (i) optimal PFS in directed networks occurs on trees, i.e. structures where one node drives the rest of the network and there is no feed-back and (ii) most directed real-world networks have large non-trivial strong components.

We argued that two ensembles of networks are relevant: (i) an ensemble of $r$-optimal networks that minimize the eigenratio but have a non-vanishing extent of the complex part of the spectrum and (ii) an ensemble of $r/c$-optimal networks whose spectra are essentially real, but which have a distinctly larger eigenratio than the $r$-optimal ensemble.

Common to both ensembles of PFS optimal networks is the expressed skewness of the out-degree distribution, which results in a typical periphery core structure. To further quantify the pecularities of the optimal networks we introduced core sets by starting from the core of high out-degree nodes and then recursively defining sets of nodes that receive only input from the previous level. Analyzing such sets reveals the existence of a hierarchy of extended cores in the optimal networks. These core sets have also been found to be in a tighly packed arrangement characterized by a small core distance. Feedback from the periphery to the core predominantly occurs from nodes in the core sets very close to the out-degree hubs. An exception to this rule is the super-hub in the $r$-optimal ensembles.

A main distinction between the ensembles is the density of double links. Whereas double links are strongly suppressed in the $r$-optimal ensemble, they are overexpressed, and this particularly in the core, in the $r/c$-optimal ensemble. While the optimized networks were not distinguished by a more than randomly expected expression ratio of any higher order motif, the frequency and arrangement of double links was identified as an important contributor to the PFS. Double links were strongly suppressed in the periphery, but were found to frequently link core nodes in the $r/c$-optimal ensemble.


\begin{thebibliography}{}
%
%
\bibitem{Arenas}
A. Arenas, A. D\'iaz-Guilera, J. Kurths, Y. Moreno, and C. Zhou, Phys. Rep.  \textbf{469}(3), 93 (2008)

\bibitem{Barabasi}
R. Albert, and A.-L. Bar\'abasi, Rev. Mod. Phys. \textbf{74}, 47 (2002)

\bibitem{Newman}
M. E. J. Newman, SIAM Rev. \textbf{45}, 167 (2003)

\bibitem{Bocc}
S. Boccaletti, V. Latora, Y. Moreno, M. Chavez, D.-U. Hwang, Phys. Rep. \textbf{424}, 175 (2006)

\bibitem{Pecora}
 L. M. Pecora and T. L. Carroll, Phys. Rev. Lett. \textbf{80}, 2109 (1998)

\bibitem{Timme}
M. Timme, Europhys. Lett. \textbf{76}, 367 (2006)

\bibitem{MB1}
M. Brede, Phys. Lett. A \textbf{372}, 5305 (2008)

\bibitem{MB2}
M. Brede, Europhys. Lett. \textbf{84}, 40004 (2008)

\bibitem{Motter1}
A. E. Motter, C. Zhou, and J. Kurths, Phys. Rev. E \textbf{71}, 016116 (2005)

\bibitem{Zhao1}
M. Zhao, T. Zhou, B.-H. Wang, G. Yan, H.-J. Yang, and W.-J. Bai, Physica A \textbf{371}, 773 (2006)

\bibitem{MB0}
M. Brede, Europ. Phys. J. B \textbf{62}, (2008) 87.

\bibitem{MB3}
M. Brede, Phys. Lett. A \textbf{372}, 2618 (2008)

\bibitem{Bernardo}
M. di Bernardo, F. Garofalo, and F. Sorrentino, Int. J. Bifurcat. Chaos \textbf{17}, 3499 (2007)

\bibitem{SW}
D. J. Watts and S. H. Strogatz, Nature \textbf{393}, 440 (1998)

\bibitem{Nishi}
T. Nishikawa, A.E. Motter, Y.-C. Lai and F.C. Hoppensteadt, Phys. Rev. Lett. \textbf{91}, 014101 (2003)

\bibitem{Hong}
H. Hong, B. J. Kim, M. Y. Choi, and H. Park, Phys. Rev. E \textbf{69}, 067105 (2004)

\bibitem{Atay}
F. M. Atay, T. Biyikoglu, and J. Jost, Physica D \textbf{224}, 35 (2006)

\bibitem{Motter2}
A. E. Motter, C. S. Zhou, and J. Kurths, Europhys. Lett. \textbf{69}, 334 (2005)

\bibitem{Hwang1}
D.-U. Hwang, M. Chavez, A. Amann, and S. Boccaletti, Phys. Rev. Lett. \textbf{94}, 138701 (2005)

\bibitem{Chavez}
M. Chavez, D.-U. Hwang, A. Amann, H. G. E. Hentschel, and S. Boccaletti, Phys. Rev. Lett. \textbf{94}, 218701 (2005)

\bibitem{Donetti}
L. Donetti, P. I. Hurtado and M. A. Mu\~noz, Phys. Rev. Lett. \textbf{95}, 188701 (2005)

\bibitem{Sole}
R. V. Sole and R. Ferrer i Cancho,in {\it Lecture Notes in Physics} \textbf{625}(Springer, Berlin, 2003)

\bibitem{Banvar}
V. Colizza, J.R. Banavar, A. Maritan, and A. Rinaldo, Phys. Rev. Lett. \textbf{92}, 198701 (2004)

\bibitem{Variano}
E. A. Variano, J. H. McCoy, and H. Lipson, Phys. Rev. Lett. \textbf{92}, 188701 (2004)

\bibitem{DN}
J. Ash and D. Newth, Physica A \textbf{380}, 673 (2007)

\bibitem{Fink}
K. S. Fink, G. Johnson, C. Carroll, D. Mar, and L. Pecora, Phys. Rev. E \textbf{61}, 5080 (2000)
\bibitem{Sommers}
H. J. Sommers, A. Crisanti, H. Somplinsky, and Y. Stein, Phys. Rev. Lett.  \textbf{60}(19), 1895(1988)
\bibitem{Moreno1}
I. Lodato, S. Boccaletti, and V. Latora, Europhys. Lett. \textbf{78}, 28001 (2007)
\bibitem{Moreno2}
Y. Moreno, M. Vazquez-Prada, A. F. Pacheco, Physica A \textbf{343}, 279 (2004)

\bibitem{Remark}
We also experimented with different cut-offs and made sure that the presented results are robust with regard to the specific choice.
\end{thebibliography}
\end{document}